\documentclass[aps,prl,twocolumn,floatfix,showpacs,superscriptaddress]{revtex4}
\usepackage[dvips]{graphicx}
\usepackage{amssymb}
\usepackage{amsmath}

\begin{document}
\title{Chaos-assisted directional light emission from microcavity lasers}

\newcommand{\MPIPKS}{\affiliation{Max-Planck-Institut f\"ur Physik
  Komplexer Systeme, N\"othnitzer Stra\ss e 38, D-01187 Dresden,
  Germany}}
\newcommand{\OPU}{\affiliation{Department of Communication
    Engineering, Okayama Prefectural University, 111 Kuboki, Soja,
    Okayama 719-1197, Japan}}
\newcommand{\JAIST}{\affiliation{Center for Nano Materials and
Technology, Japan Advanced Institute of Science and Technology, 1-1
Asahidai, Nomi, Ishikawa 923-1292, Japan}}
\newcommand{\NTTCS}{\affiliation{NTT Communication Science
Laboratories, NTT Corporation, 2-4 Hikaridai, Seika-cho, Soraku-gun,
Kyoto 619-0237, Japan}}
\newcommand{\PURDUE}{\affiliation{Birck Nanotechnology Center,
    Department of Electrical and Computer Engineering, Purdue
    University, 1205 West State Street West Lafayette, Indiana
    47907-2057, USA}}

\author{Susumu Shinohara}\MPIPKS
\author{Takahisa Harayama}\NTTCS
\author{Takehiro Fukushima}\OPU
\author{Martina Hentschel}\MPIPKS
\author{Takahiko Sasaki}\JAIST
\author{Evgenii E. Narimanov}\PURDUE

\newcommand{\degrees}{^{\circ}}
\newcommand{\etal}{{\it et al.}}

\begin{abstract}
We study the effect of dynamical tunneling on emission from
ray-chaotic microcavities by introducing a suitably designed deformed
disk cavity.
We focus on its high quality factor modes strongly localized along a
stable periodic ray orbit confined by total internal reflection.
It is shown that dominant emission originates from the tunneling from
the periodic ray orbit to chaotic ones; the latter eventually escape
from the cavity refractively, resulting in directional emission that
is unexpected from the geometry of the periodic orbit, but fully
explained by unstable manifolds of chaotic ray dynamics.
Experimentally performing selective excitation of those modes, we
succeeded in observing the directional emission in good agreement with
theoretical prediction.
This provides decisive experimental evidence of dynamical tunneling in
a ray-chaotic microcavity.
\end{abstract}

\pacs{42.55.Sa, 05.45.Mt, 42.60.Da, 42.55.Px}

\maketitle

Dynamical billiards have served as a simple and generic model in
studies of classical and quantum chaos.
Whereas originally a dynamical billiard was introduced as an abstract
model for studying the ergodic hypothesis \cite{Kelvin}, nowadays it
has been realized as a physical system that confines microwaves,
acoustics, electrons and light \cite{Stockmann}.
As for billiards for light \cite{Nockel97,microcavities,Fukushima04},
or optical microcavities, the openness of a system intrinsic to light
confinement necessitated the studies of decaying modes, or resonances
of the Helmholtz equation, which had been less studied in the field of
quantum chaos.
Resonance characteristics such as quality factors and emission
patterns have been intensively studied in terms of classical and
quantum chaos theory \cite{reviews}.
These studies are also motivated by practical applications of optical
microcavities:
When they are applied for laser resonators, there is a demand that
modes have both high quality factors and high emission directionality.

A high quality factor mode can be associated with a ray orbit confined
by total internal reflection (TIR), whose representative example is
the whispering-gallery (WG) modes of a disk cavity.
It was demonstrated that the introduction of a slight asymmetric
deformation to a disk shape enables the existence of a WG-like mode
with emission directionality \cite{Nockel97}.
However, in general, such a mode is embedded in a variety of high
quality factor modes that do not necessarily have similar emission
directionality, implying that experimentally it is not easy to
selectively excite only desired directional modes.

For a large deformation, ray dynamics inside the cavity becomes mostly
chaotic, which causes degradation of quality factors.
However, low-loss modes start to exhibit a universal emission pattern,
which can be well explained by the unstable manifolds of an unstable
periodic orbit close to the TIR condition \cite{unstable-mani}.
The universal emission pattern can be highly directional by properly
designing an unstable manifold structure \cite{Wiersig08}.
Because low-loss modes, likely to be excited with a lasing medium,
have a similar emission pattern, one can experimentally achieve
directional emission without any settings for selective excitation.
For those low-loss modes, relatively high quality factors are
attributed to wave localization along TIR-confined unstable periodic
orbits \cite{Wiersig08}, which is, however, generally weaker than
localization along a TIR-confined WG orbit or a TIR-confined stable
periodic orbit.

In this Letter, introducing a novel largely deformed disk cavity, we
study high quality factor modes with emission directionality made
possible by a dynamical tunneling phenomenon \cite{DT, CAT}.
The key point is mostly chaotic ray dynamics having a single dominant
stable periodic orbit confined by TIR.
The high quality factors come from strong localization of the modes
along the TIR-confined stable periodic orbit.
If one adopts a one-dimensional or integrable picture based, for
instance, on the Gaussian-optic theory \cite{Tureci02}, one would
expect extremely weak emission due to TIR for those modes.
However, in reality, tunneling from the periodic orbit to neighboring
chaotic orbits takes place, where the latter eventually escape out
from the cavity refractively following unstable manifolds.
Namely, light is emitted with the assistance of chaos.
Fabricating laser diodes with this deformed disk shape with an
electrode contact patterned along the TIR-confined periodic orbit, we
experimentally perform selective excitation of the high quality factor
modes and demonstrate chaos-assisted directional emission in good
agreement with theoretical prediction.

\begin{figure}[b]
\includegraphics[width=70mm]{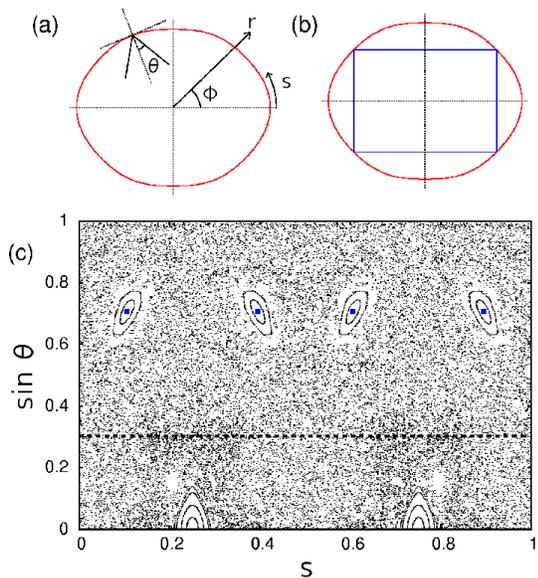}
\caption{(color online) (a) Geometry of the deformed disk cavity. (b)
The rectangular ray orbit. (c) The upper half of the Poincar\'e
surface of section for internal ray dynamics. The variable $s$ is
normalized by the total circumference. The four filled rectangles
($\blacksquare$) plotted at $\sin\theta\approx 0.7$ correspond to the
rectangular ray orbit. The critical line for total internal
reflection, $\sin\theta=1/3.3$, is plotted by a dashed line.}
\label{fig:ray-phsp}
\end{figure}

We first define the deformed disk cavity.
In the polar coordinates, its boundary is defined by
$r(\phi)=R\,(1+a\cos 2\phi+b\cos 4\phi+c\cos 6\phi)$, where $R$ is the
size parameter and $a$, $b$ and $c$ are deformation parameters fixed
as $a=0.1$, $b=0.01$ and $c=0.012$ [Fig. \ref{fig:ray-phsp}(a)].
Features of internal ray dynamics can be best described by using the
Poincar\'e surface of section (SOS).
The Poincar\'e SOS is a reduction of ray dynamics to a two-dimensional
mapping describing successive bounces of a ray orbit.
The variables of the mapping are a canonical conjugate pair called the
Birkhoff coordinates, $s$ and $\sin\theta$, where $s$ is the arclength
measured along the cavity boundary and $\sin\theta$ the tangential
momentum of an incident ray [Fig. \ref{fig:ray-phsp}(a)].
We show the SOS of our cavity in Fig. \ref{fig:ray-phsp}(c), which
reveals that ray dynamics is mostly chaotic, except for several
islands of stability.
Dominant islands are period-two islands at around $\sin\theta=0$ and 
period-four islands at around $\sin\theta=0.7$.
Figure \ref{fig:ray-phsp}(b) shows a rectangular orbit corresponding
to the period-four stable periodic points.
This orbit is confined by TIR when the laser is made from GaAs/AlGaAs
quantum-well structure (i.e., the effective refractive index $n=3.3$).
In Fig. \ref{fig:ray-phsp}(c), the critical line for TIR,
$\sin\theta=1/n$, is plotted by a dashed line, above which rays are
reflected by TIR.
The key feature of our cavity is that there exists only one dominant
stable periodic orbit above the critical line and it is surrounded by
a chaotic sea extending into the non-TIR (i.e., leaky) region,
$|\sin\theta|<1/n$.

We are interested in resonant modes corresponding to the TIR-confined
rectangular orbit.
For those modes, there are two expected emission mechanisms.
One is evanescent emission, which occurs at the four bouncing points
of the rectangular orbit, yielding emission in the tangential
direction to the cavity boundary, i.e., $\phi=45\degrees, 135\degrees,
225\degrees$, and $315\degrees$.
Another mechanism is due to dynamical tunneling
\cite{DT,CAT,Tureci02}.
An orbit on the period-four islands tunnels to a neighboring chaotic
orbit and it eventually diffuses into the leaky region, yielding
refractive light emission.
In order to capture the effect of dynamical tunneling on emission
patterns precisely, next we study wave functions of resonant modes.

Resonant modes can be obtained by solving a linear Helmholtz equation
numerically \cite{Wiersig03}.
We consider transverse electric (TE) polarization, in order to be
consistent with our experiments described below.
We systematically study resonant modes whose scaled wave number $kR$
is around $50$, where $k$ is the free-space wave number.
We checked that properties of resonant modes discussed here are
essentially the same for a larger cavity (e.g., $kR$=$100$).

\begin{figure}[b]
\includegraphics[width=50mm]{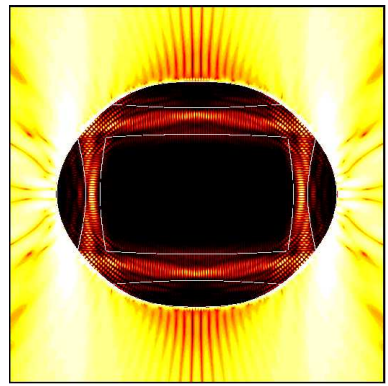}
\caption{(color online) Wave function for a rectangle-orbit mode with
  a wave number $\mbox{Re}\,kR$ $=$ $49.94$ and $\mbox{Im}\,kR$ $=$
  $-0.00012$. The intensity increases as the color changes from black
  to white. Outside the cavity, the intensity is plotted in log
  scale. The outline of the electrode contact window is indicated by
  white curves.}
\label{fig:wfunc}
\end{figure}

Employing the Husimi projection of a wave function onto the SOS, we
could classify resonant modes into three classes: rectangle-orbit
modes, WG-type modes and chaotic modes.
Here we exclude highly lossy modes such as the one residing on the
period-two islands in the leaky region.
Rectangle-orbit modes are strongly localized on the period-four
islands, while WG-type modes are strongly localized on the upper or
lower verges of the SOS (i.e., $|\sin\theta|\lessapprox 1$).
Chaotic modes are mostly distributed in the chaotic sea, even largely
spreading into the leaky region.
The first two classes of modes have high quality factors $Q\approx
10^5\sim 10^6$, as they are localized in the TIR regions, whereas
typical chaotic modes have low quality factors $Q\approx 100$.

\begin{figure}[t]
\includegraphics[width=65mm]{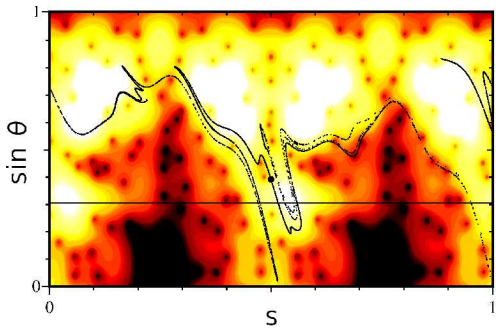}
\caption{(color online) The Husimi projection of the rectangle-orbit
  mode shown in Fig. \ref{fig:wfunc}. The intensity increases as the
  color changes from black to white. The intensity is plotted in log
  scale. The critical line for total internal reflection is plotted by
  a solid line. Unstable manifolds emanating from a period-three
  unstable periodic point at $(s,\sin\theta)=(0.5,0.389...)$,
  indicated by a filled circle, are plotted by a solid curve. The high
  intensity spots around the critical line at $s=0.04$ and $s=0.54$
  correspond to emission towards $\phi=90\degrees$ and
  $\phi=-90\degrees$, respectively.}
\label{fig:husimi}
\end{figure}

\begin{figure}[t]
\includegraphics[width=55mm]{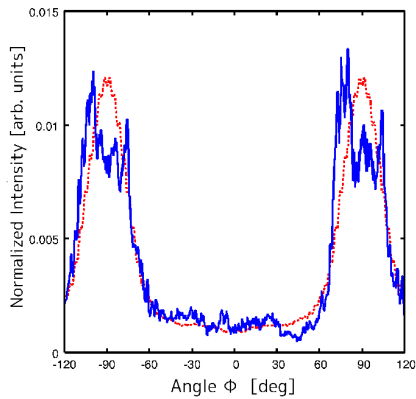}
\caption{(color online) Far-field emission patterns: experiment (solid
  curve) vs wave calculation (dotted curve). The experimental data are
  for a diode with the margin parameter $M=0.95$. The numerical data
  are for the rectangle-orbit mode shown in Fig. \ref{fig:wfunc},
  where rapid oscillation due to interference is smeared out.}
\label{fig:ffp}
\end{figure}

We plot in Fig. \ref{fig:wfunc} the wave function of a rectangle-orbit
mode, where the intensity outside the cavity is plotted in log scale
in order to clearly see the emission pattern.
Figure \ref{fig:wfunc} reveals that light is emitted towards $\phi=\pm
90^{\circ}$ on both sides of the cavity, i.e.,$\phi\approx 0\degrees$
and $180\degrees$.
By careful inspection, one finds that light is emitted from four
cavity boundary points corresponding to polar angles $\phi=\pm
4^{\circ}$ and $\phi=180\pm 4^{\circ}$.
In the corresponding far-field pattern (to be presented in
Fig. \ref{fig:ffp}), two distinct peaks appear at $\phi=\pm
90^{\circ}$ with a divergence angle $30^{\circ}$.
Moreover, in each peak, we observe the interference pattern of light
beams emitted from both sides of the cavity, whose oscillation period
is given by $\triangle\theta=360\degrees/kd$ with $d=2R\,(1+a+b+c)$
being the width of the cavity.
We confirmed that emission patterns are almost similar for all
rectangle-orbit modes.
Although it appears counterintuitive that light is emitted from
positions apart from the rectangular orbit, its mechanism can be
explained by dynamical tunneling as we see next.

Figure \ref{fig:husimi} shows the log plot of the Husimi projection of
the wave function shown in Fig. \ref{fig:wfunc}.
Because the Husimi projection is symmetric with respect to the
transformation $(s,\sin\theta)\mapsto (1-s,-\sin\theta)$, only the
upper half part is shown.
The intensity is mostly concentrated around the period-four islands,
but we can see some intensity also spread in the chaotic sea, which
can be interpreted as the effect of dynamical tunneling.
What determines the emission pattern is the intensity distribution in
the leaky region.
Around the critical line for TIR, we can observe two relatively high
intensity spots at $s=0.04$ and $s=0.54$.
It is these that are responsible for the emission towards $\phi=\pm
90\degrees$.

It is explainable by ray dynamics why emission occurs at $s=0.04$ and
$s=0.54$.
In Fig. \ref{fig:husimi}, we superimpose unstable manifolds of a
period-three unstable periodic point onto the Husimi projection.
This periodic point is located close to the critical line and thus its
unstable manifolds govern a ray-dynamical flow around the critical
line \cite{unstable-mani}.
The structure of the unstable manifolds indicate that ray orbits are
transported from the TIR region to the leaky region at around $s=0.5$
(and $s=0$ as well because of the $180\degrees$ rotational symmetry of
the cavity).
This ray-dynamical argument also explains why emission patterns of the
rectangle-orbit modes are almost similar.

\begin{figure}[t]
\includegraphics[width=68mm]{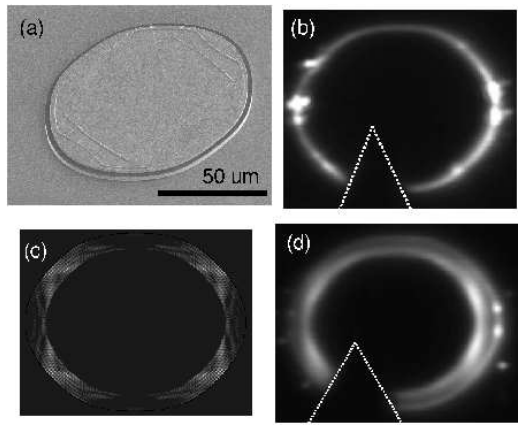}
\caption{(a) Scanning electron microscope image of a fabricated laser
  diode with the margin parameter $M=0.95$ (see text). (b) CCD photo
  of a lasing diode with $M=0.95$. (c) The wave function of a
  rectangle-orbit mode covered by the electrode metal area with
  $M=0.8$. (d) CCD photo of a lasing diode with $M=0.8$. In (b) and
  (d), the photo is taken from a slightly oblique angle and the shade
  of an electrode is indicated by dotted lines.}
\label{fig:nfp}
\end{figure}

Now we would like to examine whether the chaos-assisted light emission
can be experimentally observed.
We fabricated single-quantum-well laser diodes with the deformed disk
cavities discussed above.
The laser diodes are fabricated by applying a reactive-ion-etching
technique to a graded-index separate-confinement-heterostructure
single-quantum-well GaAs/AlGaAs structure that was grown by metal
organic chemical vapor deposition.
The lasing wavelength is around $850$ nm.
The cavity size $R$ is set as $50\,\mu$m, yielding the scaled wave
number $kR=370$.
In order to inject current only to rectangle-orbit modes, the
electrode contact window was patterned along the rectangular orbit.
Its outline is shown by white curves in Fig. \ref{fig:wfunc}.
For current injection, an electrode metal is coated on the cavity top,
which covers all over the top except for some margin of the cavity.
The precise shape of the electrode metal is defined by $r<M r(\phi)$
with the margin parameter $M (\leq 1)$.
Because of an intervening SiO$_2$ insulation layer, the electrode
metal only touches the GaAs contact layer through the contact window
\cite{Fukushima04}.
With this structure, we can selectively excite rectangle-orbit modes,
while suppressing the excitation of WG-type modes.
We fabricated samples with $M=0.8, 0.85$, and $0.95$.
A scanning electron microscope image of a fabricated laser diode with
$M=0.95$ is shown in Fig. \ref{fig:nfp}(a)

We tested the laser diodes at 25$^{\circ}$C using a pulsed current of
500-ns width at 1-kHz repetition.
Figure \ref{fig:ffp} shows a measured far-field pattern for the diode
with $M=0.95$ at the injection current 500 mA, together with a
wave-calculated far-field pattern corresponding to the mode shown in
Fig. \ref{fig:wfunc}.
Both data are normalized so that integration becomes unity.
Before the normalization of the experimental data, uniform background
contribution due to spontaneous emission is subtracted so that the
emission level around $\phi=0$ coincides with that of the
wave-calculated data.
In Fig. \ref{fig:ffp}, one can confirm excellent agreement between the
experimental and calculated data.
Far-field patterns for samples with $M=0.8$ and $0.85$ are found to be
similar to that for $M=0.95$, while substructures of the two main peaks
are different.
At the current strength 500 mA, we checked from an optical spectrum
that lasing occurs in multimode.
We expect that substructures of the main peaks observed in the
experimental data are due to the effect of multimode lasing and thus
depend on samples.
Sample-specific details will be reported elsewhere.

In order to observe which positions of the cavity light is emitted
from, we performed a near-field measurement of the cavity in lasing
operation by using a CCD camera.
The result is shown in Fig. \ref{fig:nfp}(b) for a sample with
$M=0.95$, where we can observe scattered light at the cavity boundary.
In good accordance with the numerical results, we can clearly observe
that light is emitted from two points on each side of the cavity.

Figure \ref{fig:nfp}(d) shows a near-field photo of a lasing diode
with $M=0.8$.
The pattern observed in the margin area is considered to be formed by
scattered light inside the cavity, reflecting the internal modal
pattern.
We show in Fig. \ref{fig:nfp}(c) corresponding numerical data showing
the wave function of a rectangle-orbit mode superimposed by the
$M=0.8$ electrode metal.
Comparing these results, we find bright regions of the near-field
photo corresponding to the left-arm and right-arm parts of the
rectangle-orbit mode.
This further convinces us the excitation of the rectangle-orbit modes.

In conclusion, we have demonstrated chaos-assisted directional light
emission from the deformed disk semiconductor microcavity lasers.
Theoretical analysis of resonant modes strongly localized along the
TIR-confined rectangular orbit reveals that their directional
emission, unexpected from the geometry of the rectangular orbit, can
be fully explained by tunneling to chaotic orbits eventually escaping
from the cavity following unstable manifolds around the critical line
for TIR.
In experiments, selectively exciting the rectangle-orbit modes, we
succeeded in observing the chaos-assisted directional light emission
in both the far and near fields, which provides decisive experimental
evidence of dynamical tunneling in a ray-chaotic microcavity.

We would like to thank Jan Wiersig and Jung-Wan Ryu for discussions.
S.S. and M.H. acknowledge financial support from the DFG research
group 760 ``Scattering Systems with Complex Dynamics'' and the DFG
Emmy Noether Program.

\begin{thebibliography}{99}
%
\bibitem{Kelvin} L. Kelvin, Phil. Mag. Ser. {\bf 6}, 1 (1901).
%
\bibitem{Stockmann} H.-J. St\"ockmann, {\it Quantum Chaos: An
  Introduction} (Cambridge University Press, Cambridge, England,
  1999).
%
\bibitem{Nockel97} J. U. N\"ockel and A. D. Stone, Nature (London)
  {\bf 385}, 45 (1997).
%
\bibitem{microcavities} C. Gmachl \etal, Science {\bf 280}, 1556
  (1998); S. B. Lee \etal, Phys. Rev. Lett. {\bf 88}, 033903 (2002);
  M. Lebental \etal, Appl. Phys. Lett. {\bf 88}, 031108 (2006);
  W. Fang \etal, Appl. Phys. Lett. {\bf 90}, 081108 (2007).
%
\bibitem{Fukushima04}  T. Fukushima and T. Harayama, IEEE J. Sel. Top. Quantum
  Electron. {\bf 10}, 1039 (2004).
%
\bibitem{reviews} For a review, see, for example, H. G. L. Schwefel
  \etal, in {\it Optical Microcavities}, edited by K. Vahala (World
  Scientific, Singapore, 2004).
%
\bibitem{unstable-mani} H. G. L. Schwefel \etal, J. Opt. Soc. Am. B
  {\bf 21}, 923 (2004); S. -Y. Lee \etal, Phys. Rev. A {\bf 72},
  061801 (R); S. Shinohara \etal, Phys. Rev. A {\bf 74}, 033820
  (2006); S.-B. Lee \etal, Phys. Rev. A {\bf 75}, 011802(R) (2007).
%
\bibitem{Wiersig08} J. Wiersig and M. Hentschel, Phys. Rev. Lett. {\bf
100}, 033901 (2008).
%
\bibitem{DT} M. J. Davis and E. J. Heller, J. Chem. Phys. {\bf 75},
  246 (1981).
%
\bibitem{CAT} G. Hackenbroich and J. U. N\"ockel, Europhys. Lett. {\bf
  39}, 371 (1997); V. A. Podolskiy and E. E. Narimanov,
  Opt. Lett. {\bf 30}, 474 (2005); E. E. Narimanov and
  V. A. Podolskiy, IEEE J. Sel. Top. Quantum Electron. {\bf 12}, 40
  (2006); A. B\"acker \etal, Phys. Rev. A {\bf 79}, 063804 (2009).
%
\bibitem{Tureci02} H. E. T\"ureci \etal, Opt. Express {\bf 10}, 752
  (2002).
%
\bibitem{Wiersig03} J. Wiersig, J. Opt. A: Pure Appl. Opt. {\bf 5}, 53
  (2003).
%
\end{thebibliography}
\end{document}